\documentclass[amsmath,aps,prd,showpacs,a4paper,10pt]{revtex4}

 \usepackage{epsf}
 \usepackage{graphicx}    

\begin{document}

 \newcommand{\be}[1]{\begin{equation}\label{#1}}
 \newcommand{\ee}{\end{equation}}
 \newcommand{\bea}{\begin{eqnarray}}
 \newcommand{\eea}{\end{eqnarray}}
 \def\disp{\displaystyle}

 \def\gsim{ \lower .75ex \hbox{$\sim$} \llap{\raise .27ex \hbox{$>$}} }
 \def\lsim{ \lower .75ex \hbox{$\sim$} \llap{\raise .27ex \hbox{$<$}} }

 \begin{titlepage}

 \begin{flushright}
 arXiv:0705.4002
 \end{flushright}

 \title{\Large \bf Dynamics of Quintom and Hessence Energies in
 Loop~Quantum~Cosmology}

 \author{Hao~Wei}
 \email[\,email address:\ ]{haowei@mail.tsinghua.edu.cn}
 \affiliation{Department of Physics and Tsinghua Center for
 Astrophysics,\\ Tsinghua University, Beijing 100084, China}

 \author{Shuang~Nan~Zhang}
 \affiliation{Department of Physics and Tsinghua Center for
 Astrophysics,\\ Tsinghua University, Beijing 100084, China\\
 Key Laboratory of Particle Astrophysics, Institute of High
 Energy Physics,\\
 Chinese Academy of Sciences, Beijing 100049, China\\
 Physics Department, University of Alabama in Huntsville,
 Huntsville, AL 35899, USA}

 \begin{abstract}\vspace{1cm}
 \centerline{\bf ABSTRACT}\vspace{2mm}
 In the present work, we investigate the universe dominated by
 quintom or hessence energies in Loop Quantum Cosmology~(LQC).
 Interestingly enough, we find that there are some stable
 attractors in these two cases. In the case of quintom, all
 stable attractors have the feature of decelerated expansion.
 In the case of hessence, most of stable attractors have the
 feature of decelerated expansion while one stable attractor
 can have decelerated or accelerated expansion depend on the
 model parameter. In all cases, the equation-of-state
 parameter~(EoS) of all stable attractors are larger than $-1$
 and there is no singularity in the finite future. These results
 are different from the dynamics of phantom in LQC, or the ones
 of phantom, quintom and hessence in classical Einstein gravity.
 \end{abstract}


 \maketitle

 \end{titlepage}

 \renewcommand{\baselinestretch}{1.5}



\section{Introduction}\label{sec1}
Dark energy~\cite{r1} has been one of the most active fields in
 modern cosmology since the discovery of accelerated expansion
 of our universe~\cite{r2,r3,r4,r5,r6,r7,r8}. In the observational
 cosmology of dark energy, equation-of-state parameter~(EoS)
 $w_{de}\equiv p_{de}/\rho_{de}$ plays an important role, where
 $p_{de}$ and $\rho_{de}$ are the pressure and energy density of
 dark energy respectively. Recently, evidence for $w_{de}(z)<-1$ at
 redshift $z<0.2\sim 0.3$ has been found by fitting observational
 data (see~\cite{r9,r10,r11,r12,r13,r14,r15,r16} for examples).
 In addition, many best-fits of the present value of $w_{de}$ are
 less than $-1$ in various data fittings with different
 parameterizations (see~\cite{r17} for a recent review). The present
 data seem to slightly favor an evolving dark energy with $w_{de}$
 being below $-1$ around present epoch from $w_{de}>-1$ in the near
 past~\cite{r10}. Obviously, the EoS cannot cross the so-called
 phantom divide $w_{de}=-1$ for quintessence or phantom alone.
 Some efforts have been made to build dark energy model whose EoS
 can cross the phantom divide (see
 for examples~\cite{r10,r18,r19,r20,r21,r22,r23,r24,r25,r26,r27,r28,r29,r30,r31,r32,r33,r34,r35}
 and references therein).

In~\cite{r10}, Feng, Wang and Zhang proposed a so-called quintom
 model which is a hybrid of quintessence and phantom (thus the
 name quintom). It is one of the simplest modeles whose
 EoS can cross the phantom divide. The cosmological
 evolution of the quintom dark energy was studied in~\cite{r23,r24}.
 Perturbations of the quintom dark energy were investigated
 in~\cite{r36,r37}; and it is found that the quintom model is
 stable when EoS crosses $-1$, in contrast to many dark energy
 models whose EoS can cross the phantom divide~\cite{r28}. Other
 works concerning quintom also include~\cite{r31} for examples.
 In~\cite{r18}, by a new view of quintom dark energy, one of
 us~(H.W.) and his collaborators proposed a novel non-canonical complex
 scalar field, which was named ``hessence'', to play the role of
 quintom. In the hessence model, the phantom-like role is played by
 the so-called internal motion $\dot{\theta}$, where $\theta$ is the
 internal degree of freedom of hessence. The transition from
 $w>-1$ to $w<-1$ or vice versa is also possible in the
 hessence model~\cite{r18}. The cosmological evolution of the
 hessence dark energy was studied in~\cite{r19} and then was extended
 to the more general cases in~\cite{r20}. The $w$-$w^\prime$ analysis
 of hessence dark energy was performed in~\cite{r21}. In~\cite{r30},
 the method to reconstruct hessence dark energy was proposed. We will
 briefly review the main points of quintom and hessence energies in
 Sec.~\ref{sec2}.

In fact, many works by now are considered in the framework of
 classical Einstein gravity. However, it is commonly believed that gravity
 should also be quantized, like other fundamental forces. As well-known,
 for many years, the string theory is the only promising candidate for
 quantum gravity. In the recent decade, however, the Loop Quantum
 Gravity~(LQG) (see e.g.~\cite{r38,r39,r40,r41} for reviews) has
 became a competitive alternative to the string theory. LQG is a
 leading background independent, non-perturbative approach to quantum
 gravity. At the quantum level, the classical spacetime continuum is
 replaced by a discrete quantum geometry and the operators corresponding
 to geometrical quantities have discrete eigenvalues.

 Loop Quantum Cosmology~(LQC) (see e.g.~\cite{r42,r43,r44} for
 reviews) restricts the analysis of LQG to the homogeneous and
 isotropic spacetimes. Recent investigations have shown that the
 discrete quantum dynamics can be very well approximated by an
 effective modified Friedmann dynamics~\cite{r45,r46,r47}. There
 are two types of modification to the Friedmann equation due to loop
 quantum effects~\cite{r44,r54}. The first one is based on the
 modification to the behavior of inverse scale factor operator below
 a critical scale factor $a_\ast$. So far, most of the LQC literature
 has used this one. Many interesting results have been found, for
 instance, the replacement of the classical big bang by a quantum
 bounce with desirable features~\cite{r43,r45,r48}, avoidance of many
 singularities~\cite{r49}, easier inflation~\cite{r50,r51},
 correspondence between LQC and braneworld cosmology~\cite{r52},
 and so on. However, as shown in e.g.~\cite{r53}, the first type of
 modification to Friedmann equation suffers from gauge dependence
 which can not be cured and thus lead to unphysical effects. In a
 recent paper~\cite{r64}, Magueijo and Singh provided very sharp
 results to show that it is only for the closed model that such
 modifications can be sensible and for flat models they make no
 sense.

The second type of modification to Friedmann equation is
 discovered very recently~\cite{r44,r53,r54}, which essentially
 encodes the discrete quantum geometric nature of spacetime. The
 corresponding effective modified Friedmann equation in a flat
 universe is given by~\cite{r44,r53,r54,r57,r58,r59,r60}
 \be{eq1}
 H^2=\frac{\kappa^2\rho}{3}\left(1-\frac{\rho}{\rho_c}\right),
 \ee
 where $H\equiv\dot{a}/a$ is the Hubble parameter; a dot denotes the
 derivative with respect to cosmic time $t$, and $a$ is the scale
 factor; $\rho$ is the total energy density; $\kappa^2\equiv 8\pi G$;
 and the critical density reads
 \be{eq2}
 \rho_c\equiv\frac{\sqrt{3}}{16\pi^2\gamma^3 G^2 \hbar},
 \ee
 where $\gamma$ is the dimensionless Barbero-Immirzi parameter (it is
 suggested that $\gamma\simeq 0.2375$ by the black hole thermodynamics
 in LQG~\cite{r56}). Differentiating Eq.~(\ref{eq1}) and using the
 energy conservation equation
 \be{eq3}
 \dot{\rho}+3H\left(\rho+p\right)=0,
 \ee
 one obtain the effective modified Raychaudhuri
 equation~\cite{r54,r57,r58,r59,r60}
 \be{eq4}
 \dot{H}=-\frac{\kappa^2}{2}\left(\rho+p\right)
 \left(1-2\frac{\rho}{\rho_c}\right),
 \ee
 where $p$ is the total pressure. Actually, as shown in~\cite{r64}, the
 effective modified Raychaudhuri equation can be derived directly by
 using the Hamilton's equations in LQC, without assuming the energy
 conservation equation. It is easy to check that only two of
 Eqs.~(\ref{eq1}), (\ref{eq3}) and~(\ref{eq4}) are independent of each
 other, and the third one can be derived from these two. By using the
 second type of modification to Friedmann equation, the physically
 appealing features of the first type are retained~\cite{r44}, for
 instance, resolution of big bang singularity~\cite{r44,r53}, avoidance
 of big rip and other singularities~\cite{r57,r58,r59}, inflation in
 LQC~\cite{r60}, correspondence between LQC and braneworld
 cosmology~\cite{r54}, and so on.

For the universe with a large scale factor, the first type of
 modification to the effective Friedmann equation can be neglected
 and only the second type of modification is important~\cite{r54},
 while the matter Hamiltonian ${\cal H}_M$ and the corresponding
 expressions for energy density and pressure retain the same
 classical forms~\cite{r44,r53,r54,r55,r57,r58,r59}. This is the
 particular case which we will consider here.

In the present work, we will investigate the universe dominated by
 quintom or hessence in LQC. Following~\cite{r54,r57,r58,r59}, we
 use the method of dynamical system~\cite{r61}. After a brief
 review of quintom and hessence energies in Sec.~\ref{sec2}, we
 consider the dynamics of quintom and hessence in Sec.~\ref{sec3}
 and Sec.~\ref{sec4} respectively. Interestingly enough, we find
 that there are some stable attractors in these two cases. In the
 case of quintom, all stable attractors have the feature of
 decelerated expansion. In the case of hessence, most of stable
 attractors have the feature of decelerated expansion while one
 stable attractor can have decelerated or accelerated expansion
 depend on the model parameter. In all cases, the EoS of all stable
 attractors are larger than $-1$ and there is no singularity in the
 finite future. These results are different from the dynamics of
 phantom in LQC~\cite{r58,r59}, or the ones of phantom, quintom and
 hessence in classical Einstein gravity~\cite{r63,r23,r24,r19,r20}.


\section{Quintom and hessence energies}\label{sec2}


\subsection{Quintom energy}\label{sec2a}

Phenomenologically, one may consider the Lagrangian density
 for quintom~\cite{r10,r23,r24}
 \be{eq5}
 {\cal L}_q=\frac{1}{2}\left(\partial_{\mu}\phi_1\right)^2
 -\frac{1}{2}\left(\partial_{\mu}\phi_2\right)^2-V(\phi_1,\phi_2),
 \ee
 where $\phi_1$ and $\phi_2$ are two real scalar fields and play
 the roles of quintessence and phantom respectively. Considering
 a spatially flat Friedmann-Robertson-Walker (FRW) universe and
 assuming the scalar fields $\phi_1$ and $\phi_2$ are homogeneous,
 one obtains the effective pressure and energy density for
 the quintom, i.e.
 \be{eq6}
 p=\frac{1}{2}\dot{\phi}_{1}^2-\frac{1}{2}\dot{\phi}_{2}^2
 -V(\phi_1,\phi_2),~~~~~~~
 \rho=\frac{1}{2}\dot{\phi}_{1}^2
 -\frac{1}{2}\dot{\phi}_{2}^2+V(\phi_1,\phi_2),
 \ee
 respectively. The corresponding effective EoS is given by
 \be{eq7}
 w\equiv\frac{p}{\rho}=\frac{\dot{\phi}_{1}^2
 -\dot{\phi}_{2}^2-2V(\phi_1,\phi_2)}{\dot{\phi}_{1}^2
 -\dot{\phi}_{2}^2+2V(\phi_1,\phi_2)}.
 \ee
 It is easy to see that $w\geq-1$ when
 $\dot{\phi}_{1}^2\geq\dot{\phi}_{2}^2$ while $w<-1$
 when $\dot{\phi}_{1}^2<\dot{\phi}_{2}^2$. We consider the simplest
 quintom whose $V(\phi_1,\phi_2)=V_1(\phi_1)+V_2(\phi_2)$ in the
 present work. In this case, the equations of motion for $\phi_1$
 and $\phi_2$ are given by
 \bea
 &&\ddot{\phi}_1+3H\dot{\phi}_1+\frac{dV_1}{d\phi_1}=0,\label{eq8}\\
 &&\ddot{\phi}_2+3H\dot{\phi}_2-\frac{dV_2}{d\phi_2}=0.\label{eq9}
 \eea


\subsection{Hessence energy}\label{sec2b}

Following~\cite{r18,r19,r30}, we consider a non-canonical complex
 scalar field, namely the hessence,
 \be{eq10}
 \Phi=\phi_1+i\phi_2,
 \ee
 with a Lagrangian density
 \be{eq11}
 {\cal L}_h=\frac{1}{4}\left[\,(\partial_\mu \Phi)^2
 +(\partial_\mu\Phi^\ast)^2\,\right]-U(\Phi^2
 +\Phi^{\ast 2})=\frac{1}{2}\left[\,(\partial_\mu \phi)^2
 -\phi^2 (\partial_\mu\theta)^2\,\right]-V(\phi),
 \ee
 where we have introduced two new variables $(\phi,\theta)$ to
 describe the hessence, i.e.
 \be{eq12}
 \phi_1=\phi\cosh\theta,~~~~~~~\phi_2=\phi\sinh\theta,
 \ee
 which are defined by
 \be{eq13}
 \phi^2=\phi_{1}^2-\phi_{2}^2,~~~~~~~\coth\theta=\frac{\phi_1}{\phi_2}.
 \ee
 In fact, it is easy to see that in terms of $\phi_1$ and $\phi_2$, the
 hessence can be regarded as a special case of quintom with general
 $V(\phi_1,\phi_2)$. Considering a spatially flat FRW universe with
 scale factor $a(t)$ and assuming $\phi$ and $\theta$ are homogeneous,
 from Eq.~(\ref{eq11}) we obtain the equations of motion for $\phi$ and
 $\theta$,
 \bea
 &&\ddot{\phi}+3H\dot{\phi}+\phi\dot{\theta}^2+\frac{dV}{d\phi}=0,\label{eq14}\\
 &&\phi^2\ddot{\theta}+(2\phi\dot{\phi}+3H\phi^2)\dot{\theta}=0.\label{eq15}
 \eea
 The pressure and energy density of the hessence are
 \be{eq16}
 p=\frac{1}{2}\left(\dot{\phi}^2-\phi^2\dot{\theta}^2\right)-V(\phi), ~~~~~~~
 \rho=\frac{1}{2}\left(\dot{\phi}^2-\phi^2\dot{\theta}^2\right)+V(\phi),
 \ee
 respectively. Eq.~(\ref{eq15}) implies
 \be{eq17}
 Q=a^3 \phi^2\dot{\theta}=const.
 \ee
 which is associated with the total conserved charge within the
 physical volume due to the internal symmetry~\cite{r18,r19}.
 It turns out that
 \be{eq18}
 \dot{\theta}=\frac{Q}{a^3 \phi^2}.
 \ee
 Substituting into Eqs.~(\ref{eq14}) and (\ref{eq16}), they can be
 rewritten as
 \be{eq19}
 \ddot{\phi}+3H\dot{\phi}+\frac{Q^2}{a^6\phi^3}+\frac{dV}{d\phi}=0,
 \ee
 \be{eq20}
 p=\frac{1}{2}\dot{\phi}^2-\frac{Q^2}{2a^6\phi^2}-V(\phi),~~~~~~~
 \rho=\frac{1}{2}\dot{\phi}^2-\frac{Q^2}{2a^6 \phi^2}+V(\phi).
 \ee
 Noting that the EoS $w\equiv p/\rho\,$, it is easy to see that
 $w\geq -1$ when $\dot{\phi}^2\geq Q^2/(a^6 \phi^2)$, while
 $w<-1$ when $\dot{\phi}^2< Q^2/(a^6 \phi^2)$. We refer to
 the original papers~\cite{r18,r19,r30} for more details.


\section{Dynamics of quintom energy in LQC}\label{sec3}

In this section, we consider the universe dominated by quintom
 energy in LQC. Following~\cite{r62,r23,r24,r58,r59}, we introduce
 these five dimensionless variables
 \be{eq21}
 x_1\equiv\frac{\kappa\dot{\phi}_1}{\sqrt{6}H}\,,~~~~~~~
 x_2\equiv\frac{\kappa\dot{\phi}_2}{\sqrt{6}H}\,,~~~~~~~
 y_1\equiv\frac{\kappa\sqrt{V_1}}{\sqrt{3}H}\,,~~~~~~~
 y_2\equiv\frac{\kappa\sqrt{V_2}}{\sqrt{3}H}\,,~~~~~~~
 z\equiv\frac{\rho}{\rho_c}\,.
 \ee
 We introduce $z$ just for convenience. It is expected that $z$ is not
 independent, because of Eq.~(\ref{eq6}). In fact, in the case of the
 universe dominated by quintom energy in LQC, the effective modified
 Friedmann equation, namely Eq.~(\ref{eq1}), can be rewritten as
 \be{eq22}
 \left(x_1^2-x_2^2+y_1^2+y_2^2\right)\left(1-z\right)=1.
 \ee
 Thus, one can explicitly express $z$ in terms of $x_1$, $x_2$, $y_1$
 and $y_2$. Notice that $0\leq z\leq 1$ is required by the positiveness
 of $\rho$ and $H^2$ in Eq.~(\ref{eq1}). In addition, by using
 Eq.~(\ref{eq6}), we recast Eq.~(\ref{eq4}) as
 \be{eq23}
 -\frac{\dot{H}}{H^2}=3\left(x_1^2-x_2^2\right)\left(1-2z\right),
 \ee
 which will be used extensively. Also, Eq.~(\ref{eq7}) becomes
 \be{eq24}
 w=\frac{x_1^2-x_2^2-y_1^2-y_2^2}{x_1^2-x_2^2+y_1^2+y_2^2}.
 \ee

In this work, we consider the case of quintom with
 exponential potentials
 \be{eq25}
 V_1(\phi_1)=V_{\phi_1}e^{-\lambda_1\kappa\phi_1},~~~~~~~
 V_2(\phi_2)=V_{\phi_2}e^{-\lambda_2\kappa\phi_2},
 \ee
 where $\lambda_1$ and $\lambda_2$ are dimensionless constants.
 Without loss of generality, we choose $\lambda_1$ and $\lambda_2$
 to be positive, since we can make them positive through field
 redefinition $\phi_1\to -\phi_1$, $\phi_2\to -\phi_2$ if
 $\lambda_1$ and $\lambda_2$ are negative. By the help of
 Eqs.~(\ref{eq22})---(\ref{eq25}), (\ref{eq6}) and~(\ref{eq3}),
 the equations of motion for $\phi_1$ and $\phi_2$,
 namely Eqs.~(\ref{eq8}) and~(\ref{eq9}), can be rewritten
 as an autonomous system
 \bea
 &&x_1^\prime=3x_1\left(A-1\right)
 +\sqrt{\frac{3}{2}}\lambda_1 y_1^2,\label{eq26}\\
 &&x_2^\prime=3x_2\left(A-1\right)
 -\sqrt{\frac{3}{2}}\lambda_2 y_2^2,\label{eq27}\\
 &&y_1^\prime=3y_1\left(A-\frac{\lambda_1}
 {\sqrt{6}}x_1\right),\label{eq28}\\
 &&y_2^\prime=3y_2\left(A-\frac{\lambda_2}
 {\sqrt{6}}x_2\right),\label{eq29}\\
 &&z^\prime=-6z\left(x_1^2-x_2^2\right)
 \left(1-z\right),\label{eq30}
 \eea
 where a prime denotes the derivative with respect to the so-called
 $e$-folding time $N\equiv\ln a$, and
 \be{eq31}
 A\equiv\left(x_1^2-x_2^2\right)\left(1-2z\right)
 =\left(x_1^2-x_2^2\right)\left[\,2\left(x_1^2-x_2^2
 +y_1^2+y_2^2\right)^{-1}-1\right],
 \ee
 in which we have used Eq.~(\ref{eq22}). We can obtain the critical
 points $(\bar{x}_1,\bar{x}_2,\bar{y}_1,\bar{y}_2,\bar{z})$ of the
 autonomous system Eqs.~(\ref{eq26})---(\ref{eq30}) by
 imposing the conditions $\bar{x}_1^\prime=\bar{x}_2^\prime=
 \bar{y}_1^\prime=\bar{y}_2^\prime=\bar{z}^\prime=0$. Of course,
 they are subject to the Friedmann constraint Eq.~(\ref{eq22}),
 namely $\left(\bar{x}_1^2-\bar{x}_2^2+\bar{y}_1^2
 +\bar{y}_2^2\right)\left(1-\bar{z}\right)=1$. We present the critical
 points and their existence conditions in Table~\ref{tab1}. In fact,
 there are other four critical points with
 $$\bar{x}_1=\frac{\lambda_1\lambda_2^2}{\sqrt{6}\left(\lambda_2^2
 -\lambda_1^2\right)}\,,~~~~
 \bar{x}_2=\frac{\lambda_1^2\lambda_2}{\sqrt{6}\left(\lambda_2^2
 -\lambda_1^2\right)}\,,~~~~\bar{y}_1=\pm\lambda_2\sqrt{r_q}\,,~~~~
 \bar{y}_2=\pm\lambda_1\sqrt{-r_q}\,,~~~~\bar{z}=0\,,$$
 where
 $$r_q\equiv\frac{6\lambda_2^2-\lambda_1^2\left(6+\lambda_2^2\right)}
 {6\left(\lambda_2^2-\lambda_1^2\right)^2}\,.$$
 However, the existence of $\bar{y}_1$ and $\bar{y}_2$ requires
 $r_q=0$. In this case, these four critical points reduce to
 Points~(Q2p) or~(Q4p).

 \begin{table}[htbp]
 \begin{center}
 \begin{tabular}{l|c|c}
 \hline\hline Label & \ Critical Point
 $(\bar{x}_1,\bar{x}_2,\bar{y}_1,\bar{y}_2,\bar{z})$ & \ Existence \\ \hline
 Q1p & \ $\bar{x}_1^2\geq 1$, \ $\sqrt{\bar{x}_1^2-1}$, \ 0, \ 0, \ 0 & $\bar{x}_1^2\geq 1$ \\
 Q1m & \ $\bar{x}_1^2\geq 1$, \ $-\sqrt{\bar{x}_1^2-1}$, \ 0, \ 0, \ 0 & $\bar{x}_1^2\geq 1$ \\
 Q2p & \ $\frac{\sqrt{6}}{\lambda_1}$, \ $\sqrt{\frac{6}{\lambda_1^2}-1}$, \ 0, \ 0, \ 0 & $\lambda_1\leq\sqrt{6}$ \\
 Q2m & \ $\frac{\sqrt{6}}{\lambda_1}$, \ $-\sqrt{\frac{6}{\lambda_1^2}-1}$, \ 0, \ 0, \ 0 & $\lambda_1\leq\sqrt{6}$ \\
 Q3 & \ $\frac{\lambda_1}{\sqrt{6}}$, \ 0, \ $\pm\sqrt{1-\frac{\lambda_1^2}{6}}$, \ 0, \ 0 & $\lambda_1\leq\sqrt{6}$ \\
 Q4p & \ $\sqrt{1+\frac{6}{\lambda_2^2}}$, \ $\frac{\sqrt{6}}{\lambda_2}$, \ 0, \ 0, \ 0 & always \\
 Q4m & \ $-\sqrt{1+\frac{6}{\lambda_2^2}}$, \ $\frac{\sqrt{6}}{\lambda_2}$, \ 0, \ 0, \ 0 & always \\
 Q5 & \ 0, \ $-\frac{\lambda_2}{\sqrt{6}}$, \ 0, \ $\pm\sqrt{1+\frac{\lambda_2^2}{6}}$, \ 0 & always \\
 \hline\hline
 \end{tabular}
 \end{center}
 \caption{\label{tab1} Critical points for the
 autonomous system Eqs.~(\ref{eq26})---(\ref{eq30}) and their
 existence conditions.}
 \end{table}

To study the stability of the critical points for the autonomous
 system Eqs.~(\ref{eq26})---(\ref{eq30}), we substitute linear
 perturbations $x_1\to\bar{x}_1+\delta x_1$,
 $x_2\to\bar{x}_2+\delta x_2$, $y_1\to\bar{y}_1+\delta y_1$,
 $y_2\to\bar{y}_2+\delta y_2$, and $z\to\bar{z}+\delta z$ about
 the critical point $(\bar{x}_1,\bar{x}_2,\bar{y}_1,\bar{y}_2,\bar{z})$
 into the autonomous system Eqs.~(\ref{eq26})---(\ref{eq30}) and
 linearize them. Because of the Friedmann constraint~(\ref{eq22}),
 there are only four independent evolution equations, i.e.
 \bea
 &&\delta x_1^\prime=3\bar{x}_1\delta A+3\left(\bar{A}-1\right)
 \delta x_1+\sqrt{6}\lambda_1\bar{y}_1\delta y_1,\label{eq32}\\
 &&\delta x_2^\prime=3\bar{x}_2\delta A+3\left(\bar{A}-1\right)
 \delta x_2-\sqrt{6}\lambda_2\bar{y}_2\delta y_2,\label{eq33}\\
 &&\delta y_1=3\bar{y}_1\left(\delta A-\frac{\lambda_1}{\sqrt{6}}
 \delta x_1\right)+3\left(\bar{A}-\frac{\lambda_1}{\sqrt{6}}
 \bar{x}_1\right)\delta y_1,\label{eq34}\\
 &&\delta y_2=3\bar{y}_2\left(\delta A-\frac{\lambda_2}{\sqrt{6}}
 \delta x_2\right)+3\left(\bar{A}-\frac{\lambda_2}{\sqrt{6}}
 \bar{x}_2\right)\delta y_2,\label{eq35}
 \eea
 where
 \bea
 &\bar{A}=&\left(\bar{x}_1^2-\bar{x}_2^2\right)
 \left[\,2\left(\bar{x}_1^2-\bar{x}_2^2+\bar{y}_1^2
 +\bar{y}_2^2\right)^{-1}-1\right],\label{eq36}\\
 &\delta A=&-4\left(\bar{x}_1^2-\bar{x}_2^2\right)
 \left(\bar{x}_1^2-\bar{x}_2^2+\bar{y}_1^2
 +\bar{y}_2^2\right)^{-2}\left(\bar{x}_1\delta x_1
 -\bar{x}_2\delta x_2+\bar{y}_1\delta y_1+\bar{y}_2
 \delta y_2\right)\nonumber\\
 &&+2\left[\,2\left(\bar{x}_1^2-\bar{x}_2^2
 +\bar{y}_1^2+\bar{y}_2^2\right)^{-1}-1\right]
 \left(\bar{x}_1\delta x_1-\bar{x}_2\delta x_2\right).\label{eq37}
 \eea
 The four eigenvalues of the coefficient matrix of
 Eqs.~(\ref{eq32})---(\ref{eq35}) determine the stability of
 the critical point. We present the corresponding eigenvalues
 for the critical points in Table~\ref{tab2}. It is easy to see
 that Points~(Q1m), (Q2m), (Q3), (Q4m) and~(Q5) are unstable.
 Point~(Q1p) exists and is stable under conditions $\bar{x}_1>1$,
 $\lambda_1\bar{x}_1\geq\sqrt{6}$ and
 $\lambda_2\sqrt{\bar{x}_1^2-1}\geq\sqrt{6}$. Point~(Q2p) exists
 and is stable under conditions $\lambda_1\leq\sqrt{6}$ and
 $\lambda_2\sqrt{9\lambda_1^{-2}-3/2}\geq 3$. Point~(Q4p) exists
 and is stable under condition
 $\lambda_1\sqrt{9\lambda_2^{-2}+3/2}\geq 3$.

The three stable attractors~(Q1p), (Q2p) and~(Q4p) have the common
 features $\bar{y}_1=\bar{y}_2=\bar{z}=0$ and
 $\bar{x}_1^2-\bar{x}_2^2=1$. From Eq.~(\ref{eq24}), we get the
 EoS $w=1$, which implies that the quintom behave as a stiff fluid.
 From Eq.~(\ref{eq23}), $-\dot{H}/H^2=3$. Then, we find that
 $H=t^{-1}/3$ (the integral constant can be set to zero by
 redefining the time). Thus, we obtain $a\propto t^{1/3}$. From
 Eq.~(\ref{eq3}) and $w=1$, we find that
 $\rho\propto a^{-6}\propto t^{-2}$. The universe undergoes
 decelerated expansion and there is no singularity in the
 finite future.

 \begin{table}[htbp]
 \begin{center}
 \begin{tabular}{l|c}
 \hline\hline Point & Eigenvalues \\ \hline
 Q1p & \ $-6$, \ 0, \ $3-\sqrt{\frac{3}{2}}\lambda_1\bar{x}_1$,
 \ $3-\lambda_2\sqrt{\frac{3}{2}\left(\bar{x}_1^2-1\right)}$ \\
 Q1m & \ $-6$, \ 0, \ $3-\sqrt{\frac{3}{2}}\lambda_1\bar{x}_1$,
 \ $3+\lambda_2\sqrt{\frac{3}{2}\left(\bar{x}_1^2-1\right)}$ \\
 Q2p & \ $-6$, \ 0, \ 0,
 \ $3-\lambda_2\sqrt{9\lambda_1^{-2}-\frac{3}{2}}$ \\
 Q2m & \ $-6$, \ 0, \ 0,
 \ $3+\lambda_2\sqrt{9\lambda_1^{-2}-\frac{3}{2}}$ \\
 Q3 & \ $-\lambda_1^2$, \ $\frac{\lambda_1^2}{2}$,
 \ $\frac{1}{2}\left(\lambda_1^2-6\right)$,
 \ $\frac{1}{2}\left(\lambda_1^2-6\right)$ \\
 Q4p & \ $-6$, \ 0, \ 0,
 \ $3-\lambda_1\sqrt{9\lambda_2^{-2}+\frac{3}{2}}$ \\
 Q4m & \ $-6$, \ 0, \ 0,
 \ $3+\lambda_1\sqrt{9\lambda_2^{-2}+\frac{3}{2}}$ \\
 Q5 & \ $-\frac{\lambda_2^2}{2}$, \ $\lambda_2^2$,
 \ $-\frac{1}{2}\left(\lambda_2^2+6\right)$,
 \ $-\frac{1}{2}\left(\lambda_2^2+6\right)$ \\
 \hline\hline
 \end{tabular}
 \end{center}
 \caption{\label{tab2} The corresponding eigenvalues
 for the critical points of the autonomous system
 Eqs.~(\ref{eq26})---(\ref{eq30}).}
 \end{table}


\section{Dynamics of hessence energy in LQC}\label{sec4}

In this section, we consider the universe dominated by hessence
 energy in LQC. Similar to the case of quintom,
 following~\cite{r62,r19,r20,r58,r59}, we introduce
 these five dimensionless variables
 \be{eq38}
 x\equiv\frac{\kappa\dot{\phi}}{\sqrt{6}H}\,,~~~~~~~
 y\equiv\frac{\kappa\sqrt{V}}{\sqrt{3}H}\,,~~~~~~~
 z\equiv\frac{\rho}{\rho_c}\,,~~~~~~~
 u\equiv\frac{\sqrt{6}}{\kappa\phi}\,,~~~~~~~
 v\equiv\frac{\kappa}{\sqrt{6}H}\frac{Q}{a^3\phi}\,.
 \ee
 Again, we introduce $z$ just for convenience, since it is expected
 that $z$ is not independent due to Eq.~(\ref{eq20}). In fact, in
 the case of the universe dominated by hessence energy in LQC, the
 effective modified Friedmann equation, namely Eq.~(\ref{eq1}), can
 be rewritten as
 \be{eq39}
 \left(x^2+y^2-v^2\right)\left(1-z\right)=1,
 \ee
 which can be used to explicitly express $z$ in terms of $x$, $y$ and
 $v$. Notice that $0\leq z\leq 1$ is required by the positiveness
 of $\rho$ and $H^2$ in Eq.~(\ref{eq1}). By using Eq.~(\ref{eq20}),
 the effective modified Raychaudhuri equation, namely
 Eq.~(\ref{eq4}), can be rewritten as
 \be{eq40}
 -\frac{\dot{H}}{H^2}=3\left(x^2-v^2\right)\left(1-2z\right).
 \ee
 From Eq.~(\ref{eq20}), the EoS of hessence is given by
 \be{eq41}
 w\equiv\frac{p}{\rho}=\frac{x^2-v^2-y^2}{x^2-v^2+y^2}.
 \ee

In this work, we consider the case of hessence with
 exponential potential
 \be{eq42}
 V(\phi)=V_\phi e^{-\lambda\kappa\phi},
 \ee
 where $\lambda$ is a dimensionless constant. Without loss of generality,
 we choose $\lambda$ to be positive, since we can make it positive
 through field redefinition $\phi\to -\phi$ if $\lambda$ is negative.
 By the help of Eqs.~(\ref{eq39})---(\ref{eq42}), (\ref{eq20})
 and~(\ref{eq3}), the equation of motion for $\phi$, namely
 Eq.~(\ref{eq19}), can be rewritten as an autonomous system
 \bea
 &&x^\prime=3x\left(B-1\right)-uv^2
 +\sqrt{\frac{3}{2}}\lambda y^2,\label{eq43}\\
 &&y^\prime=3y\left(B-\frac{\lambda}{\sqrt{6}}x\right),\label{eq44}\\
 &&z^\prime=-6z\left(x^2-v^2\right)\left(1-z\right),\label{eq45}\\
 &&u^\prime=-xu^2,\label{eq46}\\
 &&v^\prime=3v\left(B-1-\frac{1}{3}xu\right),\label{eq47}
 \eea
 where
 \be{eq48}
 B\equiv\left(x^2-v^2\right)\left(1-2z\right)=
 \left(x^2-v^2\right)\left[\,2\left(x^2+y^2-v^2\right)^{-1}-1\right],
 \ee
 in which we have used Eq.~(\ref{eq39}). We can obtain the critical
 points $(\bar{x},\bar{y},\bar{z},\bar{u},\bar{v})$ of the
 autonomous system Eqs.~(\ref{eq43})---(\ref{eq47}) by
 imposing the conditions $\bar{x}^\prime=\bar{y}^\prime=
 \bar{z}^\prime=\bar{u}^\prime=\bar{v}^\prime=0$. Of course,
 they are subject to the Friedmann constraint Eq.~(\ref{eq39}),
 namely $\left(\bar{x}^2+\bar{y}^2-\bar{v}^2\right)
 \left(1-\bar{z}\right)=1$. We present the critical
 points and their existence conditions in Table~\ref{tab3}.

 \begin{table}[htbp]
 \begin{center}
 \begin{tabular}{l|c|c}
 \hline\hline Label & \ Critical Point
 $(\bar{x},\bar{y},\bar{z},\bar{u},\bar{v})$ & \ Existence \\ \hline
 H1 & \ $\bar{x}^2\geq 1$, \ 0, \ 0, \ 0, \ $\pm\sqrt{x^2-1}$ \ & $\bar{x}^2\geq 1$ \\
 H2p & \ 1, \ 0, \ 0, \ 0, \ 0 \ & always \\
 H2m & \ $-1$, \ 0, \ 0, \ 0, \ 0 \ & always \\
 H3 & \ $\frac{\sqrt{6}}{\lambda}$, \ 0, \ 0, \ 0, \ $\pm\sqrt{\frac{6}{\lambda^2}-1}$ \ & $\lambda\leq\sqrt{6}$ \\
 H4 & \ $\frac{\lambda}{\sqrt{6}}$, \ $\pm\sqrt{1-\frac{\lambda^2}{6}}$, \ 0, \ 0, \ 0 \ & $\lambda\leq\sqrt{6}$ \\
 \hline\hline
 \end{tabular}
 \end{center}
 \caption{\label{tab3} Critical points for the
 autonomous system Eqs.~(\ref{eq43})---(\ref{eq47}) and their
 existence conditions.}
 \end{table}

To study the stability of the critical points for the autonomous
 system Eqs.~(\ref{eq43})---(\ref{eq47}), we substitute linear
 perturbations $x\to\bar{x}+\delta x$,
 $y\to\bar{y}+\delta y$, $z\to\bar{z}+\delta z$,
 $u\to\bar{u}+\delta u$, and $v\to\bar{v}+\delta v$ about
 the critical point $(\bar{x},\bar{y},\bar{z},\bar{u},\bar{v})$
 into the autonomous system Eqs.~(\ref{eq43})---(\ref{eq47}) and
 linearize them. Because of the Friedmann constraint~(\ref{eq39}),
 there are only four independent evolution equations, i.e.
 \bea
 &&\delta x^\prime=3\bar{x}\delta B+3\left(\bar{B}-1\right)\delta x
 -2\bar{u}\bar{v}\delta v-\bar{v}^2\delta u
 +\sqrt{6}\lambda\bar{y}\delta y,\label{eq49}\\
 &&\delta y^\prime=3\bar{y}\left(\delta B
 -\frac{\lambda}{\sqrt{6}}\delta x\right)
 +3\left(\bar{B}-\frac{\lambda}{\sqrt{6}}\bar{x}\right)
 \delta y,\label{eq50}\\
 &&\delta u^\prime=-2\bar{x}\bar{u}\delta u
 -\bar{u}^2\delta x,\label{eq51}\\
 &&\delta v^\prime=3\bar{v}\left[\delta B
 -\frac{1}{3}\left(\bar{x}\delta u+\bar{u}\delta x\right)\right]
 +3\left(\bar{B}-1-\frac{1}{3}\bar{x}\bar{u}\right)
 \delta v,\label{eq52}
 \eea
 where
 \bea
 &\bar{B}=&\left(\bar{x}^2-\bar{v}^2\right)\left[\,2\left(\bar{x}^2
 +\bar{y}^2-\bar{v}^2\right)^{-1}-1\right],\label{eq53}\\
 &\delta B=&-4\left(\bar{x}^2-\bar{v}^2\right)\left(\bar{x}^2
 +\bar{y}^2-\bar{v}^2\right)^{-2}\left(\bar{x}\delta x
 +\bar{y}\delta y-\bar{v}\delta v\right)\nonumber\\
 &&+2\left[\,2\left(\bar{x}^2+\bar{y}^2-\bar{v}^2\right)^{-1}
 -1\right]\left(\bar{x}\delta x-\bar{v}\delta{v}\right).\label{eq54}
 \eea
 The four eigenvalues of the coefficient matrix of
 Eqs.~(\ref{eq49})---(\ref{eq52}) determine the stability of
 the critical point. We present the corresponding eigenvalues
 for the critical points in Table~\ref{tab4}. It is easy to see that
 Point~(H1) exists and is stable under conditions $\bar{x}\geq 1$ and
 $\lambda\bar{x}\geq\sqrt{6}$; Point~(H2p) is stable under condition
 $\lambda\geq\sqrt{6}$; Point~(H2m) is unstable; Points~(H3) and~(H4)
 are stable when they exist under condition $\lambda\leq\sqrt{6}$.

 \begin{table}[htbp]
 \begin{center}
 \begin{tabular}{l|c}
 \hline\hline Point & Eigenvalues \\ \hline
 H1 & \ $-6$, \ 0, \ 0,
 \ $3-\sqrt{\frac{3}{2}}\lambda\bar{x}$ \\
 H2p & \ $-6$, \ 0, \ 0,
 \ $3-\sqrt{\frac{3}{2}}\lambda$ \\
 H2m & \ $-6$, \ 0, \ 0,
 \ $3+\sqrt{\frac{3}{2}}\lambda$ \\
 H3 & \ $-6$, \ 0, \ 0, \ 0 \\
 H4 & \ 0, \ $-\lambda^2$,
 \ $\frac{1}{2}\left(\lambda^2-6\right)$,
 \ $\frac{1}{2}\left(\lambda^2-6\right)$ \\
 \hline\hline
 \end{tabular}
 \end{center}
 \caption{\label{tab4} The corresponding eigenvalues
 for the critical points of the autonomous system
 Eqs.~(\ref{eq43})---(\ref{eq47}).}
 \end{table}

The stable attractors~(H1), (H2p) and~(H3) have the common features
 $\bar{y}=\bar{z}=\bar{u}=0$ and $\bar{x}^2-\bar{v}^2=1$. From
 Eq.~(\ref{eq41}), it is easy to find that the EoS $w=1$, which
 implies that the hessence behave as a stiff fluid. From
 Eq.~(\ref{eq40}), $-\dot{H}/H^2=3$. Then, we find that
 $H=t^{-1}/3$ (the integral constant can be set to zero by
 redefining the time). Thus, we obtain $a\propto t^{1/3}$. From
 Eq.~(\ref{eq3}) and $w=1$, we find that
 $\rho\propto a^{-6}\propto t^{-2}$. The universe undergoes
 decelerated expansion and there is no singularity in the
 finite future.

The stable attractor~(H4) is slightly different from other three
 stable attractors. From Eq.~(\ref{eq41}), we get the EoS
 $w=-1+\lambda^2/3\geq -1$. Note that $w\geq -1/3$ for
 $\sqrt{2}\leq\lambda\leq\sqrt{6}$, while $w<-1/3$ for
 $\lambda<\sqrt{2}$. From Eq.~(\ref{eq40}), $-\dot{H}/H^2=\lambda^2/2$.
 Then, we find that $H=2t^{-1}/\lambda^2$ (the integral constant can be
 set to zero by redefining the time). Thus, we obtain
 $a\propto t^{2/\lambda^2}$. From Eq.~(\ref{eq3}) and $w=-1+\lambda^2/3$,
 we find that $\rho\propto a^{-\lambda^2}\propto t^{-2}$. The universe
 experiences decelerated expansion for
 $\sqrt{2}\leq\lambda\leq\sqrt{6}$, or accelerated expansion for
 $\lambda<\sqrt{2}$. However, the universe cannot undergo
 super-accelerated expansion~($\dot{H}>0$) for any $\lambda$.
 Therefore, there is no singularity in the finite future for any case.


\section{Conclusion}\label{sec5}

In the framework of classical Einstein gravity, the dynamics of
 phantom, quintom and hessence have been studied in
 literature~\cite{r63,r23,r24,r19,r20}. In the case of phantom,
 the universe will end in a big rip singularity~\cite{r63}. In
 the case of quintom without direct couping between $\phi_1$
 and $\phi_2$~\cite{r23}, or with a special interaction between
 $\phi_1$ and $\phi_2$ through
 $V_{int}\sim\left[V_1\left(\phi_1\right)
 V_2\left(\phi_2\right)\right]^{1/2}$~\cite{r24},
 the phantom-dominated solution is the unique attractor and
 the big rip is inevitable. In the case of hessence~\cite{r19,r20},
 however, the big rip can be avoided, while its attractors are
 slightly different from the ones of the present work.

In the framework of LQC, the dynamics of phantom has also been
 studied~\cite{r58,r59}. It is found that there is no stable attractor
 in this case. Therefore, the phase trajectory is very sensitive to
 initial conditions. However, the big rip can be avoided and the
 universe finally enters oscillatory regime. This is mainly due to
 the quantum correction to Friedmann equation.

In the present work, we investigate the universe dominated by
 quintom or hessence energies in Loop Quantum Cosmology~(LQC).
 Interestingly enough, we find that there are some stable
 attractors in these two cases. In the case of quintom, all
 stable attractors have the feature of decelerated expansion.
 In the case of hessence, most of stable attractors have the
 feature of decelerated expansion while one stable attractor
 can have decelerated or accelerated expansion depend on the
 model parameter. In all cases, the equation-of-state
 parameter~(EoS) of all stable attractors are larger than $-1$
 and there is no singularity in the finite future. These results
 are different from the dynamics of phantom in LQC, or the ones
 of phantom, quintom and hessence in classical Einstein gravity.


\section*{ACKNOWLEDGMENTS}
We are grateful to Professor Rong-Gen~Cai for helpful discussions.
 We also thank Xin~Zhang, Zong-Kuan~Guo, Hui~Li, and Sumin~Tang,
 Shi-Chao~Tang, Jian~Hu, Yue~Shen, Xin~Liu, Lin~Lin, Jing~Jin,
 Wei-Ke~Xiao, Feng-Yun~Rao, Nan~Liang, Rong-Jia~Yang, Jian~Wang,
 Yuan~Liu for kind help and discussions. We acknowledge partial
 funding support from China Postdoctoral Science Foundation, and
 by the Ministry of Education of China, Directional Research Project
 of the Chinese Academy of Sciences under project No.~KJCX2-YW-T03,
 and by the National Natural Science Foundation of China under
 project No.~10521001.



\begin{thebibliography}{99}

\bibitem{r1}
P.~J.~E.~Peebles and B.~Ratra,
 Rev.\ Mod.\ Phys.\  {\bf 75}, 559 (2003) [astro-ph/0207347];\\
T.~Padmanabhan, Phys.\ Rept.\  {\bf 380}, 235 (2003) [hep-th/0212290];\\
S.~M.~Carroll, astro-ph/0310342;\\
R.~Bean, S.~Carroll and M.~Trodden, astro-ph/0510059;\\
V.~Sahni and A.~A.~Starobinsky,
 Int.\ J.\ Mod.\ Phys.\ D {\bf 9}, 373 (2000) [astro-ph/9904398];\\
S.~M.~Carroll, Living Rev.\ Rel.\  {\bf 4}, 1 (2001) [astro-ph/0004075];\\
T.~Padmanabhan, Curr.\ Sci.\  {\bf 88}, 1057 (2005) [astro-ph/0411044];\\
S.~Weinberg, Rev.\ Mod.\ Phys.\  {\bf 61}, 1 (1989);\\
S.~Nobbenhuis,
 Found.\ Phys.\  {\bf 36}, 613 (2006) [gr-qc/0411093];\\
E.~J.~Copeland, M.~Sami and S.~Tsujikawa,
 Int.\ J.\ Mod.\ Phys.\  D {\bf 15}, 1753 (2006)
 [hep-th/0603057];\\
A.~Albrecht {\it et al.}, astro-ph/0609591;\\
R.~Trotta and R.~Bower, astro-ph/0607066.

\bibitem{r2}
A.~G.~Riess {\it et al.} [Supernova Search Team Collaboration],
 Astron.\ J.\  {\bf 116}, 1009 (1998) [astro-ph/9805201];\\
S.~Perlmutter {\it et al.} [Supernova Cosmology Project
Collaboration], Astrophys.\ J.\  {\bf 517}, 565 (1999)
[astro-ph/9812133];\\
J.~L.~Tonry {\it et al.}  [Supernova Search Team Collaboration],
 Astrophys.\ J.\  {\bf 594}, 1 (2003) [astro-ph/0305008];\\
R.~A.~Knop {\it et al.}, [Supernova Cosmology Project
 Collaboration], Astrophys.\ J.\  {\bf 598}, 102 (2003) [astro-ph/0309368];\\
A.~G.~Riess {\it et al.} [Supernova Search Team Collaboration],
 Astrophys.\ J.\  {\bf 607}, 665 (2004) [astro-ph/0402512].

\bibitem{r3}
P.~Astier {\it et al.} [SNLS Collaboration],
 Astron.\ Astrophys.\  {\bf 447}, 31 (2006) [astro-ph/0510447];\\
J.~D.~Neill {\it et al.} [SNLS Collaboration], astro-ph/0605148.

\bibitem{r4}
C.~L.~Bennett {\it et al.} [WMAP Collaboration],
 Astrophys.\ J.\ Suppl. {\bf 148}, 1 (2003) [astro-ph/0302207];\\
D.~N.~Spergel {\it et al.} [WMAP Collaboration],
 Astrophys.\ J.\ Suppl. {\bf 148} 175 (2003) [astro-ph/0302209];\\
D.~N.~Spergel {\it et al.} [WMAP Collaboration], astro-ph/0603449;\\
L.~Page {\it et al.} [WMAP Collaboration], astro-ph/0603450;\\
G.~Hinshaw {\it et al.} [WMAP Collaboration], astro-ph/0603451;\\
N.~Jarosik {\it et al.} [WMAP Collaboration], astro-ph/0603452.

\bibitem{r5}
M.~Tegmark {\it et al.} [SDSS Collaboration],
 Phys.\ Rev.\ D {\bf 69}, 103501 (2004) [astro-ph/0310723];\\
M.~Tegmark {\it et al.} [SDSS Collaboration],
 Astrophys.\ J.\  {\bf 606}, 702 (2004) [astro-ph/0310725];\\
U.~Seljak {\it et al.}, Phys.\ Rev.\ D {\bf 71}, 103515 (2005) [astro-ph/0407372];\\
J.~K.~Adelman-McCarthy {\it et al.}  [SDSS Collaboration],
 Astrophys.\ J.\ Suppl.\  {\bf 162}, 38 (2006) [astro-ph/0507711];\\
K.~Abazajian {\it et al.} [SDSS Collaboration], astro-ph/0410239;
 astro-ph/0403325; astro-ph/0305492;\\
M.~Tegmark {\it et al.} [SDSS Collaboration],
 Phys.\ Rev.\  D {\bf 74}, 123507 (2006) [astro-ph/0608632].

\bibitem{r6}
S.~W.~Allen, R.~W.~Schmidt, H.~Ebeling, A.~C.~Fabian and
 L.~van Speybroeck, Mon.\ Not.\ Roy.\ Astron.\ Soc.\  {\bf 353}, 457 (2004)
 [astro-ph/0405340].

\bibitem{r7}
A.~G.~Riess {\it et al.} [Supernova Search Team Collaboration],
 astro-ph/0611572.\\
 The numerical data of the full sample are available at\\
 http:$/\!/$braeburn.pha.jhu.edu/$^\sim$ariess/R06
 or upon request to ariess@stsci.edu

\bibitem{r8}
W.~M.~Wood-Vasey {\it et al.} [ESSENCE Collaboration],
 astro-ph/0701041;\\
G.~Miknaitis {\it et al.} [ESSENCE Collaboration],
 astro-ph/0701043.

\bibitem{r9}
D.~Huterer and A.~Cooray,
 Phys.\ Rev.\ D {\bf 71}, 023506 (2005) [astro-ph/0404062].

\bibitem{r10}
B.~Feng, X.~L.~Wang and X.~M.~Zhang,
 Phys.\ Lett.\ B {\bf 607}, 35 (2005) [astro-ph/0404224].

\bibitem{r11}
J.~Q.~Xia, G.~B.~Zhao, B.~Feng, H.~Li and X.~M.~Zhang,
 Phys.\ Rev.\ D {\bf 73}, 063521 (2006) [astro-ph/0511625];\\
J.~Q.~Xia, G.~B.~Zhao, B.~Feng and X.~M.~Zhang,
 JCAP {\bf 0609}, 015 (2006) [astro-ph/0603393];\\
G.~B.~Zhao, J.~Q.~Xia, B.~Feng and X.~M.~Zhang, astro-ph/0603621;\\
J.~Q.~Xia, G.~B.~Zhao, H.~Li, B.~Feng and X.~M.~Zhang,
 Phys.\ Rev.\ D {\bf 74}, 083521 (2006) [astro-ph/0605366];\\
J.~Q.~Xia, G.~B.~Zhao and X.~M.~Zhang,
 Phys.\ Rev.\  D {\bf 75}, 103505 (2007) [astro-ph/0609463];\\
G.~B.~Zhao, J.~Q.~Xia, H.~Li, C.~Tao, J.~M.~Virey, Z.~H.~Zhu and
 X.~M.~Zhang, Phys.\ Lett.\  B {\bf 648}, 8 (2007) [astro-ph/0612728].

\bibitem{r12}
Y.~Wang and M.~Tegmark,
 Phys.\ Rev.\ D {\bf 71}, 103513 (2005) [astro-ph/0501351].

\bibitem{r13}
U.~Alam, V.~Sahni and A.~A.~Starobinsky,
 JCAP {\bf 0406}, 008 (2004) [astro-ph/0403687].

\bibitem{r14}
B.~A.~Bassett, P.~S.~Corasaniti and M.~Kunz,
 Astrophys.\ J.\  {\bf 617}, L1 (2004) [astro-ph/0407364];\\
A.~Cabre, E.~Gaztanaga, M.~Manera, P.~Fosalba and F.~Castander,
 Mon.\ Not.\ Roy.\ Astron.\ Soc.\ Lett.\  {\bf 372}, L23 (2006)
  [astro-ph/0603690].

\bibitem{r15}
S.~Nesseris and L.~Perivolaropoulos,
 Phys.\ Rev.\ D {\bf 70}, 043531 (2004) [astro-ph/0401556];\\
R.~Lazkoz, S.~Nesseris and L.~Perivolaropoulos,
 JCAP {\bf 0511}, 010 (2005) [astro-ph/0503230].

\bibitem{r16}
Y.~Wang and P.~Mukherjee,
 Astrophys.\ J.\  {\bf 650}, 1 (2006) [astro-ph/0604051].

\bibitem{r17}
A.~Upadhye, M.~Ishak and P.~J.~Steinhardt,
 Phys.\ Rev.\ D {\bf 72}, 063501 (2005) [astro-ph/0411803].

\bibitem{r18}
H.~Wei, R.~G.~Cai and D.~F.~Zeng,
 Class.\ Quant.\ Grav.\  {\bf 22}, 3189 (2005) [hep-th/0501160].

\bibitem{r19}
H.~Wei and R.~G.~Cai,
 Phys.\ Rev.\ D {\bf 72}, 123507 (2005) [astro-ph/0509328].

\bibitem{r20}
M.~Alimohammadi and H.~Mohseni Sadjadi,
 Phys.\ Rev.\ D {\bf 73}, 083527 (2006) [hep-th/0602268].

\bibitem{r21}
W.~Zhao and Y.~Zhang,
 Phys.\ Rev.\ D {\bf 73}, 123509 (2006) [astro-ph/0604460].

\bibitem{r22}
H.~Wei and R.~G.~Cai,
 Phys.\ Lett.\ B {\bf 634}, 9 (2006) [astro-ph/0512018];\\
H.~Wei and R.~G.~Cai,
 Phys.\ Rev.\ D {\bf 73}, 083002 (2006) [astro-ph/0603052].

\bibitem{r23}
Z.~K.~Guo, Y.~S.~Piao, X.~M.~Zhang and Y.~Z.~Zhang,
 Phys.\ Lett.\ B {\bf 608}, 177 (2005) [astro-ph/0410654].

\bibitem{r24}
X.~F.~Zhang, H.~Li, Y.~S.~Piao and X.~M.~Zhang,
 Mod.\ Phys.\ Lett.\ A {\bf 21}, 231 (2006) [astro-ph/0501652].

\bibitem{r25}
X.~Zhang,
 Int.\ J.\ Mod.\ Phys.\ D {\bf 14}, 1597 (2005)
 [astro-ph/0504586];\\
Z.~Chang, F.~Q.~Wu and X.~Zhang,
 Phys.\ Lett.\ B {\bf 633}, 14 (2006) [astro-ph/0509531];\\
X.~Zhang and F.~Q.~Wu,
 Phys.\ Rev.\ D {\bf 72}, 043524 (2005) [astro-ph/0506310];\\
X.~Zhang,
 Phys.\ Rev.\  D {\bf 74}, 103505 (2006) [astro-ph/0609699].

\bibitem{r26}
H.~S.~Zhang and Z.~H.~Zhu,
 Phys.\ Rev.\ D {\bf 73}, 043518 (2006) [astro-ph/0509895];\\
H.~S.~Zhang and Z.~H.~Zhu,
 Phys.\ Rev.\  D {\bf 75}, 023510 (2007) [astro-ph/0611834];\\
H.~S.~Zhang and Z.~H.~Zhu, astro-ph/0703245;\\
H.~S.~Zhang and Z.~H.~Zhu, arXiv:0704.3121 [astro-ph].

\bibitem{r27}
H.~Wei and R.~G.~Cai, astro-ph/0607064.

\bibitem{r28}
A.~Vikman, Phys.\ Rev.\ D {\bf 71}, 023515 (2005)
[astro-ph/0407107].

\bibitem{r29}
H.~Wei and S.~N.~Zhang,
 Phys.\ Lett.\  B {\bf 644}, 7 (2007) [astro-ph/0609597];\\
H.~Wei and S.~N.~Zhang, arXiv:0704.3330 [astro-ph].

\bibitem{r30}
H.~Wei, N.~N.~Tang and S.~N.~Zhang,
 Phys.\ Rev.\  D {\bf 75}, 043009 (2007) [astro-ph/0612746].

\bibitem{r31}
Z.~K.~Guo, Y.~S.~Piao, X.~M.~Zhang and Y.~Z.~Zhang,
 Phys.\ Rev.\  D {\bf 74}, 127304 (2006) [astro-ph/0608165];\\
M.~Z.~Li, B.~Feng and X.~M.~Zhang,
 JCAP {\bf 0512}, 002 (2005) [hep-ph/0503268];\\
X.~F.~Zhang and T.~T.~Qiu,
 Phys.\ Lett.\  B {\bf 642}, 187 (2006) [astro-ph/0603824];\\
Y.~F.~Cai, H.~Li, Y.~S.~Piao and X.~M.~Zhang,
 Phys.\ Lett.\  B {\bf 646}, 141 (2007) [gr-qc/0609039];\\
Y.~F.~Cai, M.~Z.~Li, J.~X.~Lu, Y.~S.~Piao, T.~T.~Qiu
 and X.~M.~Zhang, hep-th/0701016;\\
Y.~F.~Cai, T.~T.~Qiu, Y.~S.~Piao, M.~Z.~Li and X.~M.~Zhang,
 arXiv:0704.1090 [gr-qc];\\
R.~Lazkoz, G.~Leon and I.~Quiros,
 Phys.\ Lett.\  B {\bf 649}, 103 (2007) [astro-ph/0701353];\\
R.~Lazkoz and G.~Leon,
 Phys.\ Lett.\  B {\bf 638}, 303 (2006) [astro-ph/0602590];\\
M.~R.~Setare,
 Phys.\ Lett.\  B {\bf 641}, 130 (2006) [hep-th/0611165];\\
M.~Alimohammadi and H.~M.~Sadjadi,
 Phys.\ Lett.\  B {\bf 648}, 113 (2007) [gr-qc/0608016];\\
H.~Mohseni Sadjadi and M.~Alimohammadi,
 Phys.\ Rev.\  D {\bf 74}, 043506 (2006) [gr-qc/0605143];\\
W.~Wang, Y.~X.~Gui and Y.~Shao,
 Chin.\ Phys.\ Lett.\  {\bf 23}, 762 (2006);\\
P.~X.~Wu and H.~W.~Yu,
 Int.\ J.\ Mod.\ Phys.\  D {\bf 14}, 1873 (2005) [gr-qc/0509036].

\bibitem{r32}
P.~S.~Apostolopoulos and N.~Tetradis,
 Phys.\ Rev.\ D {\bf 74}, 064021 (2006) [hep-th/0604014].

\bibitem{r33}
E.~Elizalde, S.~Nojiri and S.~D.~Odintsov,
 Phys.\ Rev.\ D {\bf 70}, 043539 (2004) [hep-th/0405034];\\
S.~Nojiri, S.~D.~Odintsov and S.~Tsujikawa,
 Phys.\ Rev.\ D {\bf 71}, 063004 (2005) [hep-th/0501025];\\
S.~Nojiri and S.~D.~Odintsov,
 Gen.\ Rel.\ Grav.\  {\bf 38}, 1285 (2006) [hep-th/0506212];\\
S.~Capozziello, S.~Nojiri and S.~D.~Odintsov,
 Phys.\ Lett.\ B {\bf 632}, 597 (2006) [hep-th/0507182];\\
S.~Nojiri and S.~D.~Odintsov,
 Phys.\ Rev.\ D {\bf 72}, 023003 (2005) [hep-th/0505215];\\
E.~Elizalde, S.~Nojiri, S.~D.~Odintsov and P.~Wang,
 Phys.\ Rev.\ D {\bf 71}, 103504 (2005) [hep-th/0502082].

\bibitem{r34}
E.~O.~Kahya and V.~K.~Onemli, gr-qc/0612026;\\
T.~Brunier, V.~K.~Onemli and R.~P.~Woodard,
 Class.\ Quant.\ Grav.\  {\bf 22}, 59 (2005) [gr-qc/0408080].

\bibitem{r35}
I.~Y.~Aref'eva, A.~S.~Koshelev and S.~Y.~Vernov,
 Phys.\ Rev.\  D {\bf 72}, 064017 (2005) [astro-ph/0507067];\\
I.~Y.~Aref'eva and A.~S.~Koshelev,
 JHEP {\bf 0702}, 041 (2007) [hep-th/0605085];\\
I.~Y.~Aref'eva,
 AIP Conf.\ Proc.\  {\bf 826}, 301 (2006)
 [astro-ph/0410443];\\
I.~Y.~Aref'eva, L.~V.~Joukovskaya and S.~Y.~Vernov,
 hep-th/0701184.

\bibitem{r36}
G.~B.~Zhao, J.~Q.~Xia, M.~Li, B.~Feng and X.~M.~Zhang,
 Phys.\ Rev.\ D {\bf 72}, 123515 (2005) [astro-ph/0507482].

\bibitem{r37}
M.~Kunz and D.~Sapone,
 Phys.\ Rev.\  D {\bf 74}, 123503 (2006) [astro-ph/0609040].

\bibitem{r38}
C.~Rovelli,
 Living Rev.\ Rel.\  {\bf 1}, 1 (1998) [gr-qc/9710008];\\
T.~Thiemann,
 Lect.\ Notes Phys.\  {\bf 631}, 41 (2003) [gr-qc/0210094];\\
M.~Bojowald, gr-qc/0505057;\\
A.~Corichi,
 J.\ Phys.\ Conf.\ Ser.\  {\bf 24}, 1 (2005)
 [gr-qc/0507038];\\
A.~Perez, gr-qc/0409061.

\bibitem{r39}
A.~Ashtekar and J.~Lewandowski,
 Class.\ Quant.\ Grav.\  {\bf 21}, R53 (2004)
 [gr-qc/0404018];\\
A.~Ashtekar, arXiv:0705.2222 [gr-qc].

\bibitem{r40}
C.~Rovelli, {\it Quantum Gravity},
 Cambridge University Press, Cambridge (2004).

\bibitem{r41}
A.~Ashtekar,
 New J.\ Phys.\  {\bf 7}, 198 (2005) [gr-qc/0410054];\\
T.~Thiemann, hep-th/0608210.

\bibitem{r42}
M.~Bojowald,
 Living Rev.\ Rel.\  {\bf 8}, 11 (2005) [gr-qc/0601085];\\
M.~Bojowald, gr-qc/0505057.

\bibitem{r43}
A.~Ashtekar, M.~Bojowald and J.~Lewandowski,
 Adv.\ Theor.\ Math.\ Phys.\  {\bf 7}, 233 (2003)
 [gr-qc/0304074];\\
A.~Ashtekar, gr-qc/0702030.

\bibitem{r44}
A.~Ashtekar,
 AIP Conf.\ Proc.\  {\bf 861}, 3 (2006) [gr-qc/0605011].

\bibitem{r45}
A.~Ashtekar, T.~Pawlowski and P.~Singh,
 Phys.\ Rev.\ Lett.\  {\bf 96}, 141301 (2006) [gr-qc/0602086].

\bibitem{r46}
M.~Bojowald, P.~Singh and A.~Skirzewski,
 Phys.\ Rev.\  D {\bf 70}, 124022 (2004) [gr-qc/0408094].

\bibitem{r47}
P.~Singh and K.~Vandersloot,
 Phys.\ Rev.\  D {\bf 72}, 084004 (2005) [gr-qc/0507029].

\bibitem{r48}
A.~Ashtekar, T.~Pawlowski and P.~Singh,
 Phys.\ Rev.\  D {\bf 73}, 124038 (2006) [gr-qc/0604013].

\bibitem{r49}
P.~Singh and A.~Toporensky,
 Phys.\ Rev.\  D {\bf 69}, 104008 (2004) [gr-qc/0312110];\\
G.~V.~Vereshchagin, JCAP {\bf 0407}, 013 (2004) [gr-qc/0406108];\\
G.~Date and G.~M.~Hossain,
 Phys.\ Rev.\ Lett.\  {\bf 94}, 011302 (2005) [gr-qc/0407074].

\bibitem{r50}
M.~Bojowald,
 Phys.\ Rev.\ Lett.\  {\bf 89}, 261301 (2002)
 [gr-qc/0206054];\\
M.~Bojowald and K.~Vandersloot,
 Phys.\ Rev.\  D {\bf 67}, 124023 (2003) [gr-qc/0303072];\\
M.~Bojowald, J.~E.~Lidsey, D.~J.~Mulryne, P.~Singh and R.~Tavakol,
 Phys.\ Rev.\  D {\bf 70}, 043530 (2004) [gr-qc/0403106].

\bibitem{r51}
S.~Tsujikawa, P.~Singh and R.~Maartens,
 Class.\ Quant.\ Grav.\  {\bf 21}, 5767 (2004)
 [astro-ph/0311015];\\
J.~E.~Lidsey, D.~J.~Mulryne, N.~J.~Nunes and R.~Tavakol,
 Phys.\ Rev.\  D {\bf 70}, 063521 (2004) [gr-qc/0406042];\\
D.~J.~Mulryne, N.~J.~Nunes, R.~Tavakol and J.~E.~Lidsey,
 Int.\ J.\ Mod.\ Phys.\  A {\bf 20}, 2347 (2005)
 [gr-qc/0411125];\\
N.~J.~Nunes,
 Phys.\ Rev.\  D {\bf 72}, 103510 (2005) [astro-ph/0507683].

\bibitem{r52}
E.~J.~Copeland, J.~E.~Lidsey and S.~Mizuno,
 Phys.\ Rev.\  D {\bf 73}, 043503 (2006) [gr-qc/0510022].

\bibitem{r53}
A.~Ashtekar, T.~Pawlowski and P.~Singh,
 Phys.\ Rev.\  D {\bf 74}, 084003 (2006) [gr-qc/0607039].

\bibitem{r54}
P.~Singh,
 Phys.\ Rev.\  D {\bf 73}, 063508 (2006) [gr-qc/0603043].

\bibitem{r55}
P.~Singh,
 Class.\ Quant.\ Grav.\  {\bf 22}, 4203 (2005) [gr-qc/0502086].

\bibitem{r56}
A.~Ashtekar, J.~Baez, A.~Corichi and K.~Krasnov,
 Phys.\ Rev.\ Lett.\  {\bf 80}, 904 (1998) [gr-qc/9710007];\\
M.~Domagala and J.~Lewandowski,
 Class.\ Quant.\ Grav.\  {\bf 21}, 5233 (2004)
 [gr-qc/0407051];\\
K.~A.~Meissner,
 Class.\ Quant.\ Grav.\  {\bf 21}, 5245 (2004) [gr-qc/0407052].

\bibitem{r57}
M.~Sami, P.~Singh and S.~Tsujikawa,
 Phys.\ Rev.\  D {\bf 74}, 043514 (2006) [gr-qc/0605113].

\bibitem{r58}
D.~Samart and B.~Gumjudpai, arXiv:0704.3414 [gr-qc].

\bibitem{r59}
T.~Naskar and J.~Ward, arXiv:0704.3606 [gr-qc].

\bibitem{r60}
X.~Zhang and Y.~Ling, arXiv:0705.2656 [gr-qc].

\bibitem{r61}
A.~A.~Coley, gr-qc/9910074;\\
J.~Wainwright and G.~F.~R.~Ellis, {\it Dynamical Systems in Cosmology},
 Cambridge University Press, Cambridge (1997);\\
A.~A.~Coley, {\it Dynamical Systems and Cosmology}, in Series:
 Astrophysics and Space Science Library, Vol.~291, Springer (2004).

\bibitem{r62}
E.~J.~Copeland, A.~R.~Liddle and D.~Wands,
 Phys.\ Rev.\  D {\bf 57}, 4686 (1998) [gr-qc/9711068].

\bibitem{r63}
R.~R.~Caldwell,
 Phys.\ Lett.\  B {\bf 545}, 23 (2002) [astro-ph/9908168];\\
R.~R.~Caldwell, M.~Kamionkowski and N.~N.~Weinberg,
 Phys.\ Rev.\ Lett.\  {\bf 91}, 071301 (2003) [astro-ph/0302506].

\bibitem{r64}
J.~Magueijo and P.~Singh, astro-ph/0703566.


\end{thebibliography}
\end{document}